\UseRawInputEncoding
\documentclass[showpacs,twocolumn,superscriptaddress,nofootinbib]{revtex4-1} 

\usepackage{hyperref}
\hypersetup{colorlinks=true,linkcolor=blue,citecolor=blue}
\usepackage{tikz}
\usepackage{graphicx}
\usepackage{xcolor, soul}
\sethlcolor{green}
\usepackage{dcolumn}
\usepackage{bm}
\usepackage{color}
\usepackage{enumitem}
\usepackage{mathrsfs}
\usepackage{amsmath}
\usepackage{amssymb}
\usepackage{cleveref}
\usepackage{orcidlink}

\crefname{equation}{Eqn.}{Eqns.}
\crefname{figure}{Fig.}{Figs.}
\crefname{section}{Sec.}{Sec.}
\crefname{table}{Table}{Tables}
\usepackage{physics}
\providecommand{\dif}{\mathrm{d}} \def\d{\dif}
\setstcolor{red}

\newcommand{\beq}{\begin{equation}}
\newcommand{\eeq}{\end{equation}}
\newcommand{\bea}{\begin{eqnarray}}
\newcommand{\eea}{\end{eqnarray}}

\providecommand{\dif}{\mathrm{d}} % upright differential

\providecommand{\dif}{\mathrm{d}} \def\d{\dif}

\def\RS{\Sigma}

\def\d{\dif}

\graphicspath{{pic/}}

\begin{document}

\title{Testing alternative spacetimes by high-frequency quasi-periodic oscillations observed in microquasars and active galactic nuclei}

\author{Misbah Shahzadi
\orcidlink{0000-0002-3130-1602}}
\email{misbahshahzadi51@gmail.com}
\affiliation{Department of Mathematics, COMSATS University Islamabad, Lahore Campus, 54000 Lahore, Pakistan}

\author{Martin Kolo{\v s}
\orcidlink{0000-0002-4900-5537}}
\email{martin.kolos@physics.slu.cz}
\affiliation{Research Centre for Theoretical Physics and Astrophysics, Institute of Physics, \\Silesian University in Opava, Bezru\v{c}ovo n\'{a}m.13, CZ-74601 Opava, Czech Republic}

\author{Rabia Saleem
\orcidlink{0000-0002-1124-9624}}
\email{rabiasaleem@cuilahore.edu.pk}
\affiliation{Department of Mathematics, COMSATS University Islamabad, Lahore Campus, 54000 Lahore, Pakistan}

\author{Zden{\v e}k Stuchl\'ik
\orcidlink{0000-0003-2178-3588}}
\email{zdenek.stuchlik@physics.slu.cz}
\affiliation{Research Centre for Theoretical Physics and Astrophysics, Institute of Physics, \\Silesian University in Opava, Bezru\v{c}ovo n\'{a}m.13, CZ-74601 Opava, Czech Republic}

\date{\today}
\begin{abstract}
In this article, we try to capture the influence of deviation from standard Kerr black hole spacetime on observed high-frequency quasi-periodic oscillations signal. We explore the dynamics of test particles in the field of rotating compact objects governed by the various modifications of the standard Kerr black hole spacetime and apply the model of epicyclic oscillations of Keplerian discs to the observed microquasars and active galactic nuclei high-frequency quasi-periodic oscillations data. We presented a generalized formalism for the fitting of the high-frequency quasi-periodic oscillations models so-called epicyclic resonance and relativistic precession models, under the assumption of stationary, axisymmetric, and asymptotically flat spacetimes. Recently, we have used the same set of stationary, axisymmetric, and asymptotically flat spacetimes, and estimated the restrictions of spacetime parameters with the help of hot-spot data of three flares observed at Sgr~A* by GRAVITY instrument \citep{Shahzadi-et-al:2022:EPJC:}. The aim of this work is not to test a particular theoretical model or to determine and constrain its parameters, but to map a set of well-astrophysically motivated deviations from classical Kerr black hole spacetime and demonstrate which ones provide the best fit for high-frequency quasi-periodic oscillations data and could be fruitful for future exploration.
\end{abstract}

\maketitle

\setcounter{tocdepth}{2}

\tableofcontents

\section{Introduction} \label{sec:intro}

General Relativity (GR), as the accepted theory of gravity, agrees with all observations at the Solar System scale and beyond \citep{Will:2001:LRR}. Standard GR tests in weak field limit: the bending of light \citep{Will:2015:CQGra}, the gravitational redshift, and the correction to the Mercury perihelion procession \citep{Will:2018:PhRvL} have been recently supplemented by strong gravity field tests such as direct detection of gravitation waves \citep{Abbott-et-al:2016:PhRvL} and first black hole (BH) image \citep{EHT:2019:ApJ:}. One of the important GR predictions is the existence of compact collapsed astrophysical objects - BHs - where the strong gravity regime is manifested. However, the non-linear behaviour and strong-field structure of GR still remain elusive and difficult to test \citep{Psa:2008:LRR}.

Observational data based on the dynamics of the whole Universe affirm that the major part of the Universe is invisible to direct detection. This invisible (dark) component can be divided into dark matter (DM) and dark energy (DE) \citep{Cal-Kam:2009:Natur}. Both DM and DE, which can be well represented by the cosmological constant model for example, have large relevance in astrophysical processes, see \citep{Stu-Hle:1999:PRD,Stuchlik:2005:MPLA,Sla-Stu:2005:CQGra,Bal-Mot-Now:2007:MNRAS,Sha-etal:2019:PDU:,Stu-etal:2020:Universe:}. Modern cosmological observations reveal that our Universe is composed of $68.3 \%$ DE, $26.8 \%$ DM, and $4.9 \%$ ordinary matter \citep{Plank-Coll:2014:A&A,Rez:2017:ApJ}. The DM surrounding the galaxies and clusters does not interact with the baryonic matter but can be observed by its gravitational effects on visible matter. Babcock \citep{Bab:1939:LicOB} examined the rotational speed of luminous objects in Andromeda galaxy and found that rotational speed continuously increases as one moves away from the center of these objects. This demonstrates that the outer region of that luminous part is dominated by matter which does not shine. Zwicky \citep{Zwicky:2009:GReGr} found a large amount of unseen (non-luminous) matter in the Universe rather than the seen (luminous) and detected the non-luminous mass lying outside the luminous parts of the galaxies. Besides these theoretical observations, there is no experimental success in detecting DM yet.

In addition to the need for DE and DM, in the study of BHs, a problem that appears in GR is the presence of singularities that are points or set of points where the geodesic is interrupted and the physical quantities diverge \citep{Haw-Pen:1970:PRSLSA,Bro-Rub:2013:bhce:conf}. It is believed that the problem of singularity occurs because the theory is classical and that in the quantum theory of gravity, this problem would be solved. This, together with some long-standing problems in GR (like difficulties in explaining the accelerated Universe and galaxy rotation curves, etc), has motivated the study of viable alternative theories of gravity. Such theories, also known as modified theories of gravity, aim to reproduce GR in the weak-field regime, but they can differ substantially from it in the strong curvature regime, where non-linear effects become dominant. These modified theories of gravity are developed by modifying the matter distribution or gravitational part of the Einstein-Hilbert action.

Modified theories of gravity continue to attract widespread attention among cosmologists and astrophysicists alike. Although Einstein’s theory of gravity has recognized us with explanations of physical observations, it fails in accounting for various other physical phenomena \citep{Misner:1973:grav:book:,Sha-Teu:1983:bhwd:book:,Ellis-et-al:2012:reco:book:}. This has prompted researchers to modify the classical GR or to adopt the alternative theories of gravity. The foregoing decade has witnessed a huge influx of both astrophysical and cosmological models borne out of modified gravity theories.

The astrophysical BH candidates can be classified into three major classes, depending on the mass of BHs: stellar-mass BHs having mass $M{\sim}20~M_{\odot}-100~M_{\odot}$ located in X-ray binary systems; supermassive BHs with $M{\sim}10^{5}~M_{\odot}-10^{10}~M_{\odot} $ situated in galactic nuclei; and intermediate-mass BHs having mass $M{\sim}10^{2}~M_{\odot}-10^{4}~M_{\odot}$ \citep{Narayan:2005:NJPh,Nitz-Wang:2021:PhRvL:}. The class of intermediate-mass BHs is still debatable because their observations are indirect and dynamical measurements of their masses are still lacking.

Microquasars are binary systems composed of a BH and a companion (donor) star. Matter flowing from the companion star onto the BH forms an accretion disk, and relativistic jets, which are the bipolar outflow of matter along the BH's accretion disk rotation axis. Due to friction, the matter in the accretion disk becomes hot and emits electromagnetic radiation, including X-rays in the vicinity of the BH horizon.

The quasi-periodic oscillations (QPOs) in X-ray flux light curves have long been observed in stellar-mass BH binaries and are considered one of the most efficient tests of strong gravity models. These variations appear very close to the BH, and present frequencies that scale inversely with the mass of the BH. The current technical possibilities to measure the frequencies of QPOs with high precision allow us to get useful knowledge about the central object and its background. According to the observed frequencies of QPOs, which cover the range from a few mHz up to $0.5$kHz, different types of QPOs were distinguished. Mainly, these are the high-frequency (HF) and low-frequency (LF) QPOs with frequencies up to 500~Hz and up to 30~Hz, respectively. The oscillations of HF QPOs in BH microquasars are usually stable and detected with the twin peaks which have a frequency ratio close to $3:2$ \citep{Rem-McCli:2006:ARAA:}. However, this phenomenon is not universal, the HF QPOs have been observed in only $11$ out of $7000$ observations of $22$ stellar mass BHs \citep{Bel-et-al:2012:MNRAS}. The oscillations usually occur only in specific states of hardness and luminosity, moreover, in X-ray binaries, HF QPOs occur in an ``anomalous" high-soft state or steep power law state, both corresponding to a luminous state with a soft X-ray spectrum.

One can obtain helpful information about the bounds of parameters of the system, using the methods of spectroscopy (frequency distribution of photons) and timing (photon number time dependence) for particular microquasars \citep{Rem-McCli:2006:ARAA:}. In this connection, the binary systems having BHs, being compared to neutron star systems, seem to be promising due to the reason that any astrophysical BH is assumed to be a Kerr BH (corresponding to the unique solution of GR in 4D for uncharged BHs which does not violate the weak cosmic censorship conjecture and no-hair theorem) that is specified by only two parameters: the spin parameter and BH mass.

After the first observation of QPOs, there were many efforts to fit the observed QPOs, and various models have been presented, such as the hot-spot models, warped disk models, disko-seismic models, epicyclic resonance (ER) models, relativistic precession (RP) models and many versions of resonance models \cite{Abramowicz-etal:2003:PASJ:}. The most extended is thus the so-called geodesic oscillatory model where the observed frequencies are associated with the frequencies of the geodesic orbital and epicyclic motion \citep{Stu-Kot-Tor:2013:ASTRA:}. It is interesting that the characteristic frequencies of HF QPOs are close to the values of the frequencies of the test particles, geodesic epicyclic oscillations in the regions near the innermost stable circular orbit (ISCO) which makes it reasonable to construct the model involving the frequencies of oscillations associated with the orbital motion around Kerr BHs \citep{Stu-Kot-Tor:2013:ASTRA:}. However, until now, the exact physical mechanism of the generation of HF QPOs is unknown, since none of the models can fit the observational data from different sources \citep{Bur:2005:RAG:}. Even more serious situation has been exposed in the case of HF QPOs related to accretion disks orbiting supermassive BHs in active galactic nuclei (AGNs) \citep{Kot-etal:2020:AAP:,Smi-Tan-Wag:2021:APJ:}. One possibility to overcome this problem is associated with the electromagnetic interactions of slightly charged matter moving around a magnetized BH \citep{Kol-Stu-Tur:2015:CLAQG:,Kol-Tur-Stu:2017:EPJC:,Stu-Kol-Tur:2022:PASJ:}. Here, we focus on different possibilities associated with the internal rotation of accreting matter. 

In this study, we consider the classical Kerr BHs, rotating charged BHs (Kerr-Newman (KN), braneworld, dyonic, Kerr-Taub-NUT, KN-Taub-NUT), many different bumpy spacetimes in GR (Johannsen-Psaltis, Hartle-Thorne, Kerr-Q, Quasi-Kerr, accelerating-rotating), rotating regular BHs in GR (Bardeen, Ay\'{o}n-Beato-Garica (ABG), Hayward), various metrics in several theories of gravity (Kerr-Sen BHs in heterotic string theory, Born-Infeld BHs in Einstein-Born-Infeld theory, Kalb-Ramond BHs in heterotic string theory, Gauss-Bonnet BHs in Einstein-Gauss-Bonnet theory, Konoplya-Rezzolla-Zhidenko BHs in an unknown gravity, Kerr-MOG BHs in scalar-tensor-vector theory, BHs with Weyl corrections, BHs in Rastall gravity, rotating regular BHs in conformal massive gravity, rotating regular BHs in Einstein-Yang-Mills theory, hairy BHs), as well as rotating BHs modified by DM or quintessence (dirty BHs, BHs in perfect fluid DM (PFDM), BHs in cold DM halo, BHs in scalar field DM halo, Hayward BHs in PFDM, BHs in quintessence), and explore the orbital and epicyclic motion of neutral test particles in the background of well-motivated considered rotating BHs. We look especially for the existence and properties of the harmonic or QPOs of neutral test particles. The quasi-harmonic oscillations around a stable equilibrium location and the frequencies of these oscillations are then compared with the frequencies of the HF QPOs observed in microquasars GRS 1915+105, GRO 1655-40, XTE 1550-564, and XTE J1650-500 as well as AGNs TON S 180, ESO 113-G010, 1H0419-577, RXJ 0437.4-4711, 1H0707-495, RE J1034+396, Mrk 766, ASASSN-14li, MCG-06-30-15, XMMU 134736.6+173403, Sw J164449.3+573451, MS 2254.9-3712 \citep{Sha-etal:2006:ApJ:,Rem-McCli:2006:ARAA:,Smi-Tan-Wag:2021:APJ:}.

Throughout the paper, we use the space-like signature ($-,+,+,+$) and the system of units in which $c = 1$ and $G = 1$. However, for the expressions with an astrophysical application and estimates, we use the units with the gravitational constant and the speed of light. Greek indices are taken to run from 0 to 3; Latin indices are related to the space components of the corresponding equations.

%%%%%%%%%%%%%%%%%%%%%%%%%%%%%%%%%%%%%%%%%%%%%%%%%%%%%%%%%%%%%%%%%%%%%%%%%%%
%%%%%%%%%%%%%%%%%%%%%%%%%%%%%%%%%%%%%%%%%%%%%%%%%%%%%%%%%%%%%%%%%%%%%%%%%%%
\section{Stationary and axisymmetric Spacetimes} \label{SAS}
%%%%%%%%%%%%%%%%%%%%%%%%%%%%%%%%%%%%%%%%%%%%%%%%%%%%%%%%%%%%%%%%%%%%%%%%%%%
%%%%%%%%%%%%%%%%%%%%%%%%%%%%%%%%%%%%%%%%%%%%%%%%%%%%%%%%%%%%%%%%%%%%%%%%%%%

In four-dimensional GR, the no-hair theorem \citep{Israel:1967:PhRv,Carter:1971:PhRvL} states that the uncharged rotating BHs are uniquely characterized by only two parameters, the mass $M$, and spin $a$ of the BH, and are governed by the Kerr metric. This metric is a unique axisymmetric, stationary, asymptotically flat, and vacuum solution of the Einstein field equations which possesses an event horizon, but there are no closed timelike curves in an exterior domain. Due to the weak cosmic censorship conjecture \citep{Penrose:1969:NCimR}, the central singularity is always behind the event horizon. However, the hypothesis that the astrophysical BH candidates are characterized by the Kerr spacetimes still lacks direct evidence, furthermore, the GR has been tested only in the regime of weak gravity \citep{Will:2014:LRR}. For strong gravitational fields, the GR could be broken down and astrophysical BHs might not be the Kerr BHs as predicted by the no-hair theorem \citep{Joh-Psa:2011:PhRvD}. Several parametric deviations from the Kerr metric have been proposed to investigate the observational signatures in both the electromagnetic and gravitational-wave spectra that differ from the expected Kerr signals.

The line element of an arbitrary, stationary, axisymmetric, and asymptotically flat spacetime with reflection symmetry reads
\beq
\d s^2 = g_{tt} \d{t}^2 + g_{rr} \d{r}^2 + g_{\theta\theta} \d\theta^2 + g_{\phi\phi} \d\phi^2 + 2 g_{t\phi} \d{t} \d \phi, \label{Kerr-likeMetric}
\eeq
where metric components $g_{\alpha \beta}$ are functions of $r$, $\theta$, and some additional parameters. In the following, we consider several stationary, axisymmetric, and asymptotically flat spacetimes both in GR and modified theories of gravity. The exact forms of the metrics can be found in the given references below or in review \citep{Shahzadi-et-al:2022:EPJC:}.

%%%%%%%%%%%%%%%%%%%%%%%%%%%%%%%%%%%%%%%%%%%%%%%%%%%%%%%%%%%%%%%%%%%%%%%%%%%
\subsection{Classical (Kerr) BHs in GR}
%%%%%%%%%%%%%%%%%%%%%%%%%%%%%%%%%%%%%%%%%%%%%%%%%%%%%%%%%%%%%%%%%%%%%%%%%%%

The nonzero components of the metric tensor $g_{\mu\nu}$, describing the geometry of the well-known classical neutral rotating Kerr BH can be written in the standard Boyer-Lindquist coordinates in the form \citep{Kerr:1963:PRL:,Carter:1968:PR:}
\bea
g_{tt} &=& -\left(\frac{\Delta-a^{2}\sin^{2}\theta}{\RS}\right), \quad
g_{rr} = \frac{\RS}{\Delta}, \quad g_{\theta\theta} = \RS,\nonumber\\
g_{\phi\phi} &=& \frac{\sin^{2}\theta}{\RS}\left[(r^{2}+a^{2})^{2}-\Delta
a^{2}\sin^{2}\theta\right], \nonumber\\
g_{t\phi}&=&\frac{a\sin^{2}\theta}{\RS}\left[\Delta-(r^{2}+a^{2})\right] ,
\label{MetricCoef}
\eea
with
\bea
\Delta &=& r^2 - 2Mr + a^2, \\ \RS &=& r^2 + a^2 \cos^2\theta, \label{KerrSigma}
\eea
where, $M$ and $a$ are the mass and rotation parameters of the BH, respectively. The spin parameter $a$ is bounded by $a \leq M$. The horizons for Kerr BH can be found by solving the condition $\Delta = 0$. For details of its properties, see \citep{Car:1973:BlaHol:}.

%%%%%%%%%%%%%%%%%%%%%%%%%%%%%%%%%%%%%%%%%%%%%%%%%%%%%%%%%%%%%%%%%%%%%%%%%%%
\subsection{Charged BHs in GR}
%%%%%%%%%%%%%%%%%%%%%%%%%%%%%%%%%%%%%%%%%%%%%%%%%%%%%%%%%%%%%%%%%%%%%%%%%%%

According to the no-hair theorem, BH solutions of the Einstein-Maxwell equations of GR (combining the field equations of gravity and electromagnetism) are fully characterized by their mass $M$, rotation parameter $a$, and electric charge $Q$. There are many kinds of charges such as electric, magnetic, tidal, dyonic, etc. In the following, we consider BH solutions with different charges.
\begin{itemize}
\item KN BHs \citep{Mis-Tho-Whe:1973:Gra:,Bal-Bic-Stu:1989:BAC:,Azreg-etal:2019:IJMPD:}
\item Braneworld BHs \citep{Ali-Gur:2005:PhRvD:,Khan-etal:2019:PDU:}, tested for QPOs in \citep{Stu-Kot:2009:GReGr:,Banerjee-etal:2021:JCAP:}
\item Dyonic charged BHs \citep{Kasuya:1982:PhRvD:,Stu:1983:BAC:}
\item Kerr-Taub-Nut BHs \citep{Dem-New:1966:BAPSS:,Miller:1973:JMP:}, tested for QPOs in \citep{Cha-Bha:2018:PhRvD:}
\item KN-Taub-Nut BHs \citep{Dem-New:1966:BAPSS:,Miller:1973:JMP:}
\end{itemize}

%%%%%%%%%%%%%%%%%%%%%%%%%%%%%%%%%%%%%%%%%%%%%%%%%%%%%%%%%%%%%%%%%%%%%%%%%%%
\subsection{Bumpy spacetimes in GR}
%%%%%%%%%%%%%%%%%%%%%%%%%%%%%%%%%%%%%%%%%%%%%%%%%%%%%%%%%%%%%%%%%%%%%%%%%%%

It is possible that the spacetime around massive compact objects which are assumed to be BH is not described by the Kerr metric but by a metric that can be considered as a perturbation of the Kerr metric, and is usually known as bumpy (non-Kerr) spacetime \citep{Col-Hug:2004:PhRvD:}. These spacetimes have multipoles and possess some features that deviate slightly from the Kerr spacetime, reducing to the classical Kerr BH solutions when the deviation is zero. Here, we consider some bumpy spacetimes in GR.
\begin{itemize}
\item Johannsen-Psaltis spacetime \citep{Joh-Psa:2011:PhRvD}, tested for QPOs in \citep{Cosimo:2012:JCAP:}
\item Hartle-Thorne spacetime \citep{Har-Tho:1968:ApJ:}, tested for QPOs in \citep{Gabriela-etal:2019:ApJ:,All-Ali:2021:JCAP:}
\item Kerr-Q spacetime \citep{All-Fir-Mas:2020:CQGra:}, tested for QPOs in \citep{All-Ali:2021:JCAP:}
\item Quasi-Kerr BHs \citep{Gla-Kos-Sta:2006:CQGra:}
\item Accelerating and rotating BHs \citep{Gri-Pod:2005:CQG:}
\end{itemize}

%%%%%%%%%%%%%%%%%%%%%%%%%%%%%%%%%%%%%%%%%%%%%%%%%%%%%%%%%%%%%%%%%%%%%%%%%%%
\subsection{Rotating regular BHs in GR}
%%%%%%%%%%%%%%%%%%%%%%%%%%%%%%%%%%%%%%%%%%%%%%%%%%%%%%%%%%%%%%%%%%%%%%%%%%%

The regular BHs are non-singular exact solutions of the Einstein field equations minimally coupled to non-linear electrodynamics, satisfy the weak energy condition, and yield alteration to the classical BHs \citep{Stu-Sch:2015:IJMPD}. The regular BHs are constructed to be regular everywhere, i.e., the Ricci scalar and the components of the Riemann tensor are finite $\forall ~ r \geq 0$. However, the rotating regular BH solutions have been obtained by the non-complexification procedure of the Newman-Janis algorithm and they are not exact solutions (exact up to $a^2$, where $a$ is the rotation parameter) \cite{Azreg:2014:PhRvD:}. Recently, the precise constraints on the physical charges of regular and various other BHs have been proposed with the help of Event Horizon Telescope \citep{Kocherlakota-etal:2021:PhRvD:}. There are three well-known rotating regular BHs in GR, namely
\begin{itemize}
\item Regular Bardeen BHs \citep{Bardeen:1968:Tbilisi,Tos-etal:2014:PRD:}, tested for QPOs in \citep{Ban-etal:2022J:CAP:}
\item Regular ABG BHs \citep{Ayo-Gar:1998:PhRvL,Azreg:2014:PhRvD:}
\item Regular Hayward BHs \citep{Hayward:2006:PRL:}
\end{itemize}

%%%%%%%%%%%%%%%%%%%%%%%%%%%%%%%%%%%%%%%%%%%%%%%%%%%%%%%%%%%%%%%%%%%%%%%%%%%
\subsection{Rotating BHs in alternative theories of gravity}
%%%%%%%%%%%%%%%%%%%%%%%%%%%%%%%%%%%%%%%%%%%%%%%%%%%%%%%%%%%%%%%%%%%%%%%%%%%

The late-time acceleration of the Universe is surely the most challenging problem in cosmology. Many cosmological observations indicate that the accelerated expansion of the Universe is due to the existence of a mysterious form of energy known as DE. Modern astrophysical and cosmological models are also faced with two severe theoretical problems, that can be summarized as the DM (non or weakly interacting), and the DE problem. The two observations, namely, the mass discrepancy in galactic clusters, and the behaviour of the galactic rotation curves, suggest the existence of a DM at galactic and extra-galactic scales. Recently, several modified theories of gravity have been proposed to address these two intriguing and exciting problems facing modern physics. These modified theories of gravity are constructed by modifying the gravitational or matter part of the Einstein-Hilbert action. In addition, both non-rotating and rotating BH solutions have been derived in these modified theories of gravity \citep{Sen:1992:PhRvL,Moffat:2015:EPJC:,Sha-Sha:2018:SJETP}. In the following, we consider the several rotating BH solutions in many different modified theories of gravities.
\begin{itemize}
  \item Kerr-Sen BHs \citep{Sen:1992:PhRvL}
  \item Einstein-Born-Infeld BHs \citep{Julio:2004:CQGra}
  \item Kalb-Ramond BHs \citep{Kum-Gho-Wan:2020:PhRvD,Farruh-etal:2022:EPJC:}
  \item Einstein-Gauss-Bonnet BHs \citep{Maselli-etal:2015:ApJ:,Kum-Gho:2020:JCAP,Shay-etal:2022:PhRvD:}
  \item Konoplya-Rezzolla-Zhidenko BHs \citep{Rez-Zhi:2014:PhRvD:,Kon-Zhi:2016:PhLB}
  \item Kerr-MOG BHs \citep{Moffat:2015:EPJC:,Sha-Sha:2017:EPJC:}, tested for QPOs in \citep{Kol-Sha-Stu:2020:EPJC}
%\item Kaluza-Klein BHs \citep{Larsen:2000:NuPhB:, Gha-et-al:2020:PhRvD:}
  \item BHs with Weyl corrections \citep{Chen-Jin:2014:PhRvD:}
  \item BHs in Rastall gravity \citep{Xu-et-al:2018:EPJC:}
%\item Charged Weyl BHs \citep{FAT-Oli-Vil:2021:Galax:}
  \item Regular BHs in conformal massive gravity \citep{Jus-et-al:2020:PhRvD:}
  \item Regular BHs in Einstein-Yang-Mills theory \citep{Jus-et-al:2021:PhRvD}
  \item Hairy BHs \citep{Con-Ova-Cas:2021:PhRvD:}
% \item Horava BHs \citep{Gho-Hat:2010:PhRvD:} 
\end{itemize}

%%%%%%%%%%%%%%%%%%%%%%%%%%%%%%%%%%%%%%%%%%%%%%%%%%%%%%%%%%%%%%%%%%%%%%%%%%%
\subsection{Rotating BHs modified by DM or quintessence field}
%%%%%%%%%%%%%%%%%%%%%%%%%%%%%%%%%%%%%%%%%%%%%%%%%%%%%%%%%%%%%%%%%%%%%%%%%%%

Recently, with the help of Event Horizon Telescope's observations of BH shadows, it has been proposed that the existence of BHs in the Universe is almost universally accepted \citep{EHTSC:2019:ApJ:}. Inspired by this, many physicists have begun to study the interaction between DM and BHs \citep{Kava-et-el:2020:PhRvD:, Nar-et-al:2020:PhRvD:, Xu-Wang-Tang:2021:JCAP:,Khalifeh-etal:2020:PDU:,Khalifeh:2021:MNRAS:,Pug-Stu:2022:PhRvD:,Jusufi:2023:}. Due to the existence of the supermassive BHs at the centers of galaxies, the strong gravitational potential of the BH concentrates a large number of DM particles near the BH horizon \citep{Gon-Silk:1999:PhRvL:}. The DM density increases by orders of magnitude due to the BH's gravitational field. Therefore, if DM particles can annihilate into gamma-ray radiation, the intensity of gamma-ray radiation near the BH will increase greatly, which provides a good opportunity for us to detect the DM annihilation signal. A series of DM models have been proposed in the literature, some of which we consider here.
\begin{itemize}
  \item BHs in DM (dirty BHs) \citep{Pan-Rod:2020:ChJPh:}, tested (non-rotating case) for QPOs in \citep{Stu-Vrb:2022:APJ:} 
  \item BHs in PFDM \citep{Hou-Xu-Wan:2018:JCAP,Shay-etal:2022:PhRvD:}
  \item BHs in cold DM halo \citep{Xu-et-el:2018:JCAP:}
  \item BHs in scalar field DM halo \citep{Xu-et-el:2018:JCAP:}
  \item Hayward BHs in PFDM \citep{Ma-etal:2021:MPLA:}
%\item BHs in DM spike \citep{Sou-etal:2021:ApJ:} 
%\item Deformed BHs in DM spike \citep{Xu-etal:2021:JCAP:}
\item BHs in quintessence \citep{Xu-Wang:2017:PhRvD:,Ift-Sha:2019:EPJC,She-etal:2020:PhRvD:,Mustafa-etal:2022:ChPhC:}
\end{itemize}
Note that the special model of BH spacetime modified roughly by any kind of DM was tested by HF~QPOs recently in \citep{Stu-Vrb:2021:JCAP:,Stu-Vrb:2022:APJ:}.

%-------------------------------------------------------------------------%
\begin{figure*}
\begin{center}
\includegraphics[width=\hsize]{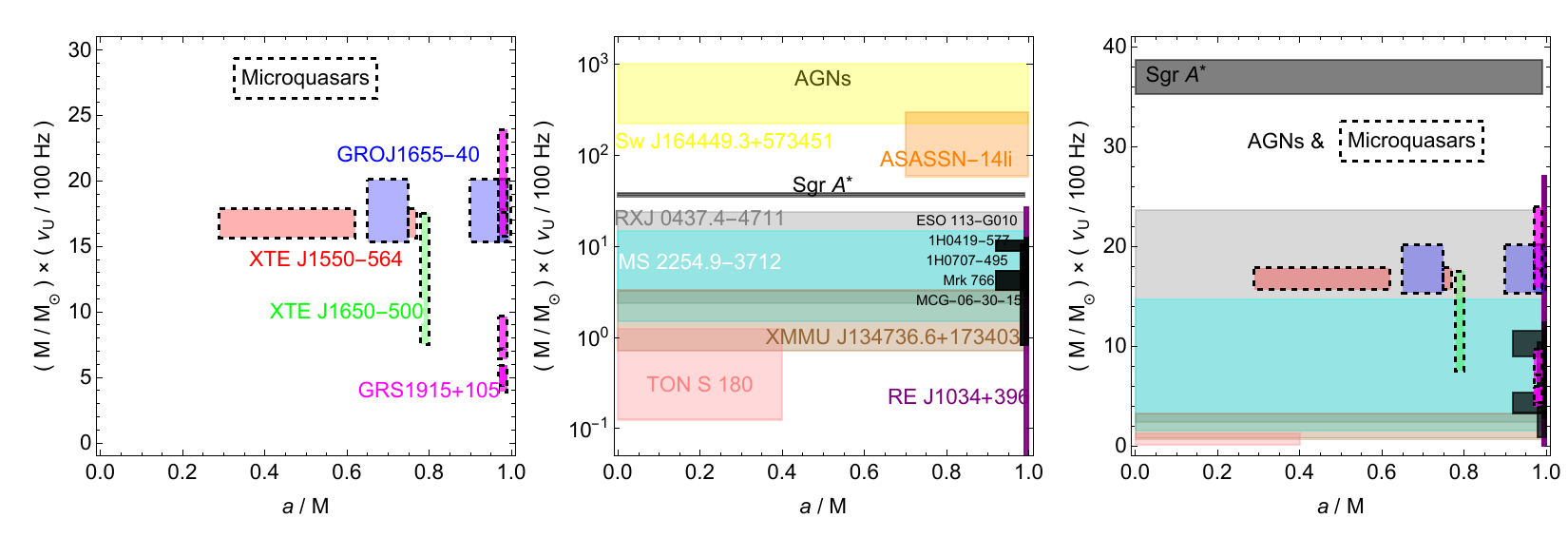}
\end{center}
%\vspace{-3pt}
\caption{The position of microquasars, AGNs, and Sagittarius (Sgr)~A* depending on the product of mass, frequencies, and spin parameter $a/M$. The shaded regions represent the objects with mass estimates, often with large errors, but no spin estimates in the literature. The colored blocks specify those objects for which the spin is estimated, while the black regions indicate those objects for which only a lower limit of spin is known.
\label{Quasars}}
\end{figure*}
%-------------------------------------------------------------------------%

%%%%%%%%%%%%%%%%%%%%%%%%%%%%%%%%%%%%%%%%%%%%%%%%%%%%%%%%%%%%%%%%%%%%%%%%%%%
\section{Quasi-periodic oscillations} 
%%%%%%%%%%%%%%%%%%%%%%%%%%%%%%%%%%%%%%%%%%%%%%%%%%%%%%%%%%%%%%%%%%%%%%%%%%%

Within this article, we will restrict our attention to test particle dynamics which can be approximated as geodesics motion. The Hamiltonian for the neutral particle motion can be written as
\beq
H=\frac{1}{2}g^{\alpha \beta} p_{\alpha} p_{\beta} + \frac{1}{2}m^2,
\eeq
where $ p^{\mu}=m u^{\mu} $ is the four-momentum of the particle with mass $m$, $u^{\alpha} = \frac{\d x^{\alpha}}{\d\tau}$ denotes the four-velocity, and $ \tau $ is the proper time. The Hamilton equations of motion can be written as
\beq
\frac{\d x^{\alpha}}{\d \zeta}\equiv m u^{\alpha} = \frac{\partial H}{\partial p_{\alpha}}, \quad
\frac{\d p_{\alpha}}{\d \zeta} = -\frac{\partial H}{\partial x^{\alpha}},
\eeq
where $ \zeta = \tau/m$ is the affine parameter. Using Hamiltonian formalism, one can find the equations of motion of particles. Due to the symmetries of the BH geometry, there exist two constants of motion, i.e., specific energy $E$, and specific angular momentum $L$, which remain conserved throughout the motion and can be expressed as
\bea
u_{t} &=& g_{tt}u^{t}+g_{t\phi}u^{\phi} = -E, \\
u_{\phi}  &=& g_{\phi\phi}u^{\phi}+g_{t\phi}u^{t} = L.
\eea
Using the above equations, we can write
\bea
\frac{\d t}{\d \tau} &=& - \frac{E g_{\phi \phi} + L g_{t \phi}}{g_{tt} g_{\phi \phi} - g_{t \phi}^2},\\\
\frac{\d \phi}{\d \tau} &=& \frac{L g_{tt} + E g_{t \phi}}{g_{tt} g_{\phi \phi} - g_{t \phi}^2}.
\eea
From the conservation of the rest mass, $g_{\alpha \beta} u^{\alpha} u^{\beta} = -1$, we have 
\beq
 \dot{r}^{2} g_{rr} +  \dot{\theta}^{2} g_{\theta \theta } = V_{\rm eff} (r, \theta, E, L),
\eeq
where the effective potential $V_{\rm eff}$ takes the form
\beq
V_{\rm eff} (r, \theta, E, L) = \frac{L^2 g_{tt} + E^2 g_{\phi \phi} + 2 E L g_{t \phi} }{- g_{tt} g_{\phi \phi} + g_{t \phi}^2} - 1.\label{veff}
\eeq
The effective potential (\ref{veff}) is very important since it enables us to demonstrate the general properties of the test particle dynamics, avoiding the necessity to solve the equations of motion.

Due to the reflection and axisymmetric properties of spacetime, for the existence of equatorial circular orbits, we have
\beq
\frac{\d r}{\d \tau} = \frac{\d \theta}{\d \tau} = \frac{\d^2 r}{\d \tau^2 } = 0.
\eeq
Consequently, the equation describing the circular motion of the particle reduces to the following form
\beq
g_{tt,r}\dot{t}^2 + 2 g_{t \phi, r}\dot{t} \dot{\phi} + g_{\phi \phi, r}\dot{\phi}^2 = 0,
\eeq
and the orbital frequency $\Omega_{\phi}$ of test particles can be written in the form
\beq
\Omega_{\phi} = \frac{\d \phi}{\d \tau} = \frac{-g_{t\phi,r} \pm \sqrt{(g_{t\phi,r})^2 - g_{tt,r} ~ g_{\phi\phi,r}}}{g_{\phi\phi,r}},\label{Freq-1}
\eeq
where the upper and lower signs refer to the prograde, and retrograde orbits, respectively. It is clear from Eq.~(\ref{Freq-1}), that the orbital frequency $\Omega_{\phi}$ in independent of the metric coefficients $g_{rr}$ and $g_{\theta \theta}$, while the partial derivatives of $g_{tt}, g_{\phi \phi}$, and $g_{tr}$ with respect to the radial coordinate $r$ are involved. 

The circular equatorial orbits are governed by the condition \citep{Kol-Stu-Tur:2015:CLAQG:,Kol-Tur-Stu:2017:EPJC:}
\beq
V_{\rm eff}(r, E, L) = 0, \quad \partial_{r} V_{\rm eff} (r,E,L) = 0.\label{Veff-1}
\eeq
The energy $E$ and angular momentum $L$ of circular orbits can be found by solving the system of Eqs. (\ref{Veff-1}), and the orbital frequency $\Omega_{\phi}$ in terms of constants of motion can be written as
\beq
\Omega_{\phi} = \frac{p_{\phi}}{p_{t}} = - \frac{g_{tt} L + g_{t \phi} E}{g_{t \phi} L + g_{\phi \phi } E}.\label{OrbFreq}
\eeq
Combining Eqs.~(\ref{Freq-1}), and (\ref{OrbFreq}), along with the condition $V_{\rm eff} (r,E,L) = 0$, one can find the energy and angular momentum in terms of orbital frequency $\Omega_{\phi}$ as
\bea
E &=&   \frac{- g_{tt} - g_{t \phi} \Omega_{\phi}}{\sqrt{ - g_{tt} - 2 g_{t \phi} \Omega_{\phi} - g_{\phi \phi} \Omega_{\phi}^{2} }},\\\
L &=& \pm \frac{g_{t \phi} + g_{\phi \phi} \Omega_{\phi}}{\sqrt{ - g_{tt} - 2 g_{t \phi} \Omega_{\phi} - g_{\phi \phi} \Omega_{\phi}^{2}}},
\eea
where the upper and lower signs correspond to the prograde and retrograde orbits, respectively. The innermost stable equatorial circular orbits so-called ISCOs are governed by the Eq.~(\ref{Veff-1}) along with the condition $\partial_{rr} V_{\rm eff} (r,E,L) = 0$. The position of ISCO is one of the parameters that are very sensitive to the value of the BH spin. 

The radial and latitudinal epicyclic frequencies can be easily calculated by considering the small perturbations around circular orbits along the radial and the latitudinal directions, respectively. If $\delta_r$ and $\delta_\theta$ are the small displacements along the mean orbit, i.e., $r = r_{0} + \delta_r$, and $\theta = \theta_{0} + \delta_{\theta}$, they lead to the following differential equations
\bea
\frac{\d^2 \delta_{r}}{\d t^2} + \Omega_{r}^{2} \delta_{r}  =  0,\\
\frac{\d^2 \delta_{\theta}}{\d t^2} + \Omega_{\theta}^{2} \delta_{\theta}  =  0,
\eea
where the radial $\Omega_{r}$ and the latitudinal $\Omega_{\theta}$ frequencies take the form
\bea\label{Freq-2}
\Omega_{r}^{2} &=&  \frac{-1}{2 g_{rr} (\dot{t})^2} \frac{\partial^{2} V_{\rm eff}}{\partial r^{2}},\\\label{Freq-3}
\Omega_{\theta}^{2} &=& \frac{-1}{2 g_{\theta \theta} (\dot{t})^2} \frac{\partial^{2} V_{\rm eff}}{\partial \theta^{2}}.
\eea
The expressions for fundamental frequencies (\ref{Freq-1}), (\ref{Freq-2}),  and (\ref{Freq-3}) are given in dimensionless form.
In physical units, one needs to extend the corresponding  formulae by the factor $c^{3}/GM$. Then, the radial, latitudinal, and orbital frequencies in $Hz$, measured by the distant observers are given by
\beq\label{nu_rel}
\nu_{\alpha}=\frac{1}{2\pi}\frac{c^{3}}{GM} \, \Omega_{\alpha}[{\rm Hz}],
\eeq
where $\alpha \in \{r,\theta,\phi\}$. The analysis of the characteristics of these frequencies and their relations are given in the following section.

%-------------------------------------------------------------------------%
\begin{figure*}
\begin{center}
\includegraphics[width=0.75\hsize]{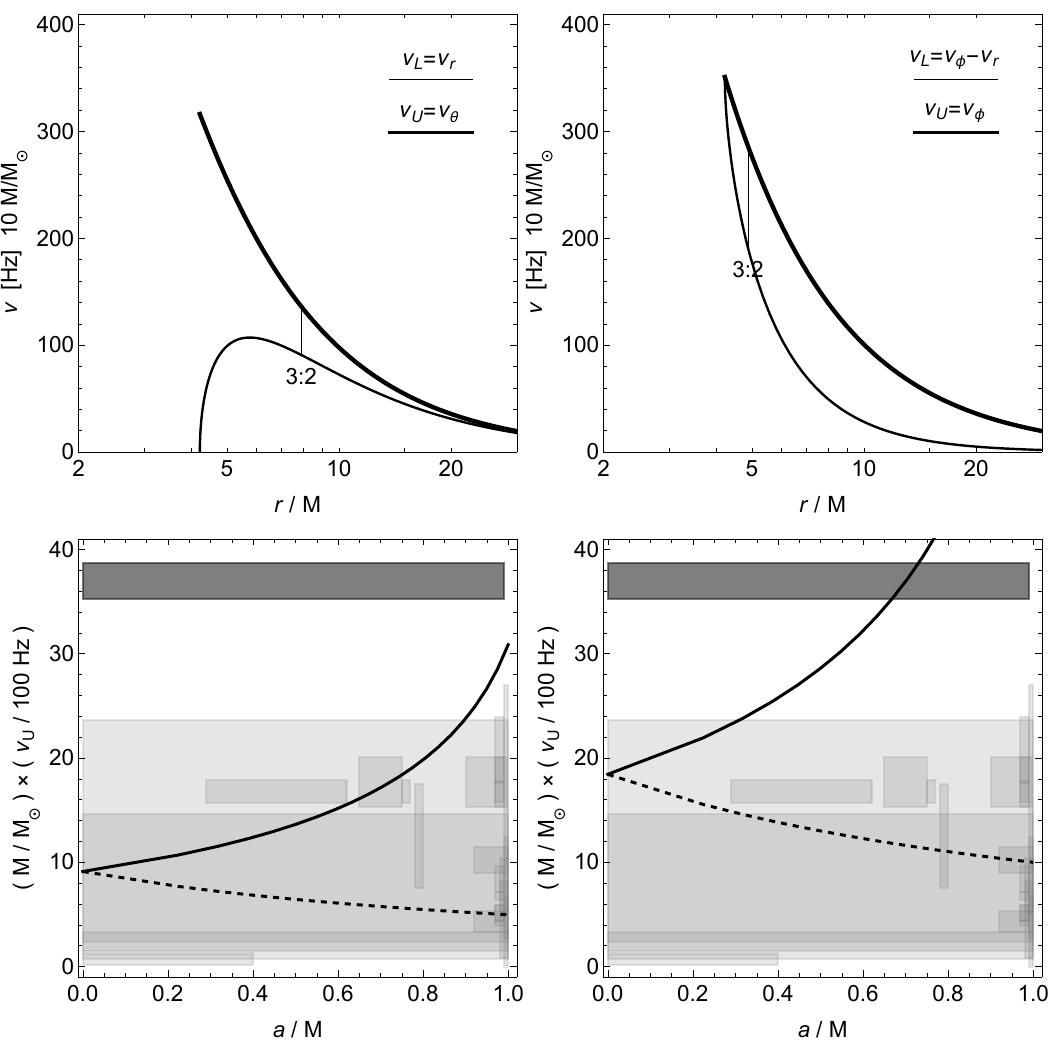}
\end{center}
\caption{First row: Radial profiles of lower $\nu_{\rm L}(r)$ and upper $\nu_{\rm U}(r)$ frequencies for ER (left) and RP (right) HF QPOs models in the background of classical Kerr BH. The BH spin $a=0.5$ has been used. The vertical lines show the position $\nu_{\rm U}:\nu_{\rm L}=3:2$ of resonant radii $r_{3:2}$. Second row: Fitting of microquasars and AGNs sources with HF QPOs ER (left) and RP (right) models. The BH spin $a/M$ is given on the horizontal axis, while on the vertical axis, we give BH mass $M$ multiplied by the observed upper HF QPOs frequency $\nu_{\rm U}$. Gray boxes show the values of BH spin $a$ and mass $M$, estimated from the observations, see Tab. \ref{tab1}. Solid curves represent the upper test particle frequency at $r_{3:2}$ resonant radii $f_{\rm U} = \nu_r(r_{3:2},a)$ as a function of BH spin $a$ for co-rotating particles, while dashed curves are plotted for contra-rotating particles.
\label{Kerr_fits}}
\end{figure*}
%-------------------------------------------------------------------------%

%%%%%%%%%%%%%%%%%%%%%%%%%%%%%%%%%%%%%%%%%%%%%%%%%%%%%%%%%%%%%%%%%%%%%%%%%%%
\begin{table*}[!ht]
\begin{center}
\begin{tabular}{c l l l l l}
\hline
Name	&	BH Spin	&	$M_\mathrm{BH}$ [$M_\odot$]	&	$f_\mathrm{QPO}$ [Hz]	&	&	Object Type \\
\hline
XTE J1550-564	&	$0.75 < a < 0.77^{r}$ 	&	$9.1^{+0.6}_{-0.6}$		&	184 	& & microquasar \\
				&	$0.29 < a < 0.62^{r,c}$ &							&	184 	\\
\hline
XTE J1650-500	&	$0.78 < a < 0.8^{r}$	&	$5.0^{+2.0}_{-2.0}$ 	&	250		& & microquasar \\
\hline
GROJ1655-40	& $0.65 < a < 0.75^{c}$ & $5.9^{+0.8}_{-0.8}$ &	300	& & microquasar \\
			& $0.9 < a < 0.998^{r}$	&					  & 300		\\
			& $0.97 < a < 0.99^{r}$	& 					  & 300	\\
\hline
GRS1915+105	& $0.97 < a < 0.99^{r,c}$ &	$9.5-14.4$ &	41	& & microquasar \\
			& 						  &			   &	67	\\
			&						  &			   &	166 \\
\hline
 Sgr~A* &  $0.4$ &	$4.3\times10^6$	&	$0.886\times10^{-3}$	& 	& Galactic center\\
\hline
 	&		&	log $M_\mathrm{BH}$	&		&	QPO Band	&	 \\
\hline
TON S 180	&	$ < 0.4$	&	$6.85^{+0.5}_{-0.5}$	&	$5.56\times10^{-6}$	&	EUV	&	NLS1 \\\\
ESO 113-G010	&	0.998	&	$6.85^{+0.15}_{-0.24}$	&	$1.24\times10^{-4}$	&	X	&	NLS1 \\\\
ESO 113-G010	&	0.998	&	$6.85^{+0.15}_{-0.24}$	&	$6.79\times10^{-5}$	&	X	& NLS1 \\\\
1H0419-577	&	$ > 0.98$	&	$8.11^{+0.50}_{-0.50}$	&	$2.0\times10^{-6}$	&	EUV	&	Sy1 \\\\
RXJ 0437.4-4711	&	--	&	$7.77^{+0.5}_{-0.5}$	&	$1.27\times10^{-5}$	&	EUV	&	Sy1 \\\\
1H0707-495	&	$>0.976$	&	$6.36^{+0.24}_{-0.06}$	&	$2.6\times10^{-4}$	&	X	&	NLS1	\\\\
RE J1034+396	&	0.998	&	$6.0^{+1.0}_{-3.49}$	&	$2.7\times10^{-4}$	&	X	&	NLS1	\\\\
Mrk 766	&	$>0.92$	&	$6.82^{+0.05}_{-0.06}$	&	$1.55\times10^{-4}$	&	X	&	NLS1	\\\\
ASASSN-14li	&	$>0.7$	&	$6.23^{+0.35}_{-0.35}$	&	$7.7\times10^{-3}$	&	X	&	TDE	\\\\
MCG-06-30-15	&	$>0.917$	&	$6.20^{+0.09}_{-0.12}$	&	$2.73\times10^{-4}$	&	X	&	NLS1	\\\\
XMMU J134736.6+173403	&	--	&	$6.99^{+0.46}_{-0.20}$	&	$1.16\times10^{-5}$	&	X		\\\\
Sw J164449.3+573451	&	--	&	$7.0^{+0.30}_{-0.35}$	&	$5.01\times10^{-3}$	&	X	&	TDE	\\\\
MS 2254.9-3712	&	--	&	$6.6^{+0.39}_{-0.60}$	&	$1.5\times10^{-4}$	&	X	&	NLS1 \\
\hline
\end{tabular}
\caption{
Observational data for QPOs around stellar-mass and supermassive BHs \citep{Smi-Tan-Wag:2021:APJ:}. In objects where multiple HF QPOs are reported in small integer ratios, only the lowest frequency is given. Superscripts on the spin ranges indicate the measurement methods: Fe~K$\alpha$ reflection spectroscopy ($r$) or continuum fitting ($c$). Restrictions on mass and spin of the BHs located in them, based on measurements independent of the HF QPO measurements given by the optical measurement for mass estimates and by the spectral continuum fitting for spin estimates \citep{Sha-etal:2006:ApJ:,Rem-McCli:2006:ARAA:}.
\label{tab1}
}
\end{center}
\end{table*}
%%%%%%%%%%%%%%%%%%%%%%%%%%%%%%%%%%%%%%%%%%%%%%%%%%%%%%%%%%%%%%%%%%%%%%%%%%%

%-------------------------------------------------------------------------%
\begin{figure*}
\centering
\includegraphics[width=\hsize]{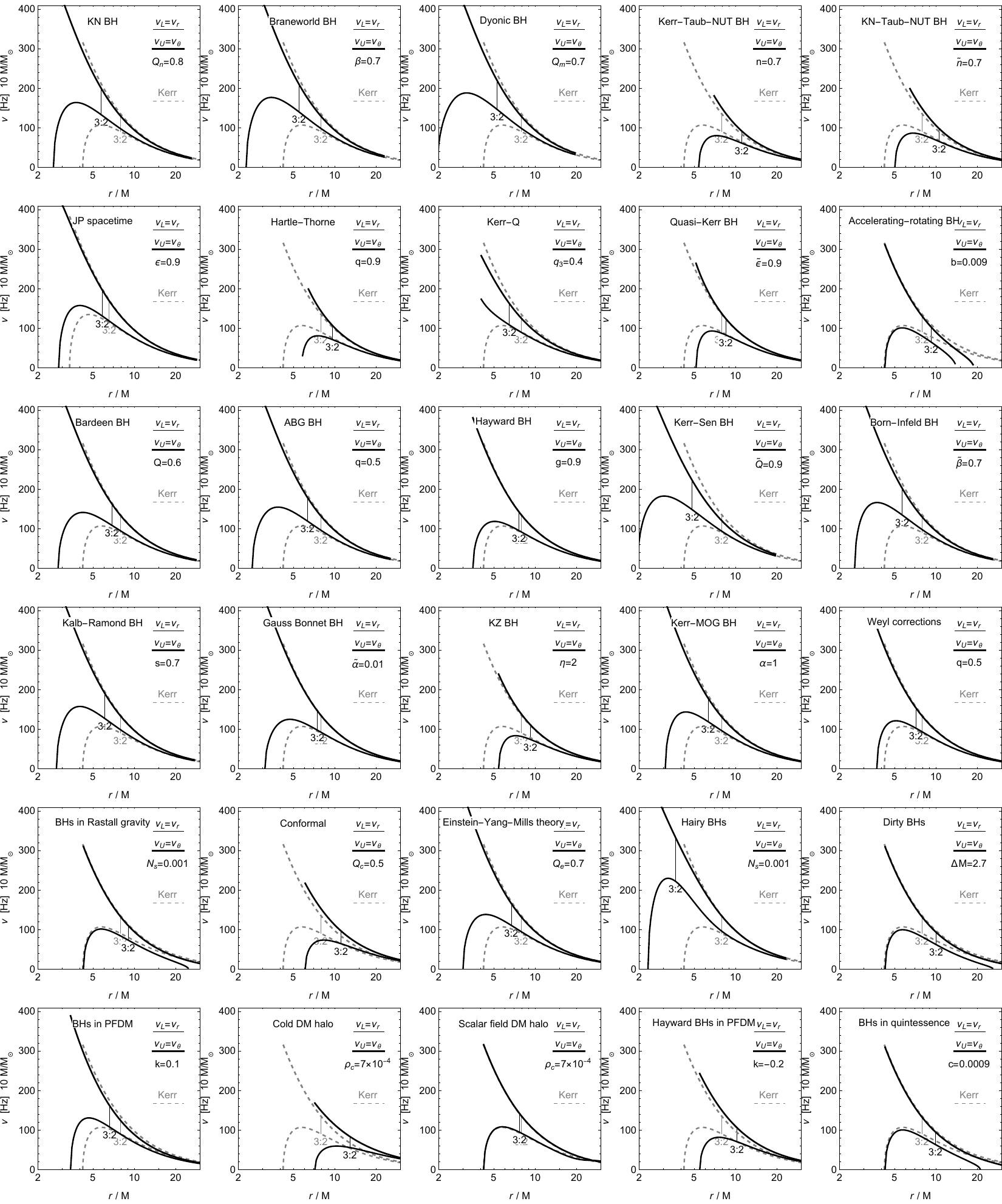}
\caption{Radial profiles of lower $\nu_{\rm L}(r)$ and upper $\nu_{\rm U}(r)$ frequencies for ER HF QPOs model in the background of many different stationary, axisymmetric and asymptotically flat spacetimes. We compare the Kerr BH limit with the well-motivated deviations from the classical Kerr BH solution. The frequencies are plotted for classical Kerr BH as gray curves, while black curves are for non-Kerr BHs. The BH spin is taken $a=0.5$ for all spacetimes. The position $\nu_{\rm U}:\nu_{\rm L}=3:2$ of resonant radii $r_{3:2}$ is also plotted.
\label{ER-32ratio}}
\end{figure*}
%-------------------------------------------------------------------------%

%-------------------------------------------------------------------------%
\begin{figure*}
\centering
\includegraphics[width=\hsize]{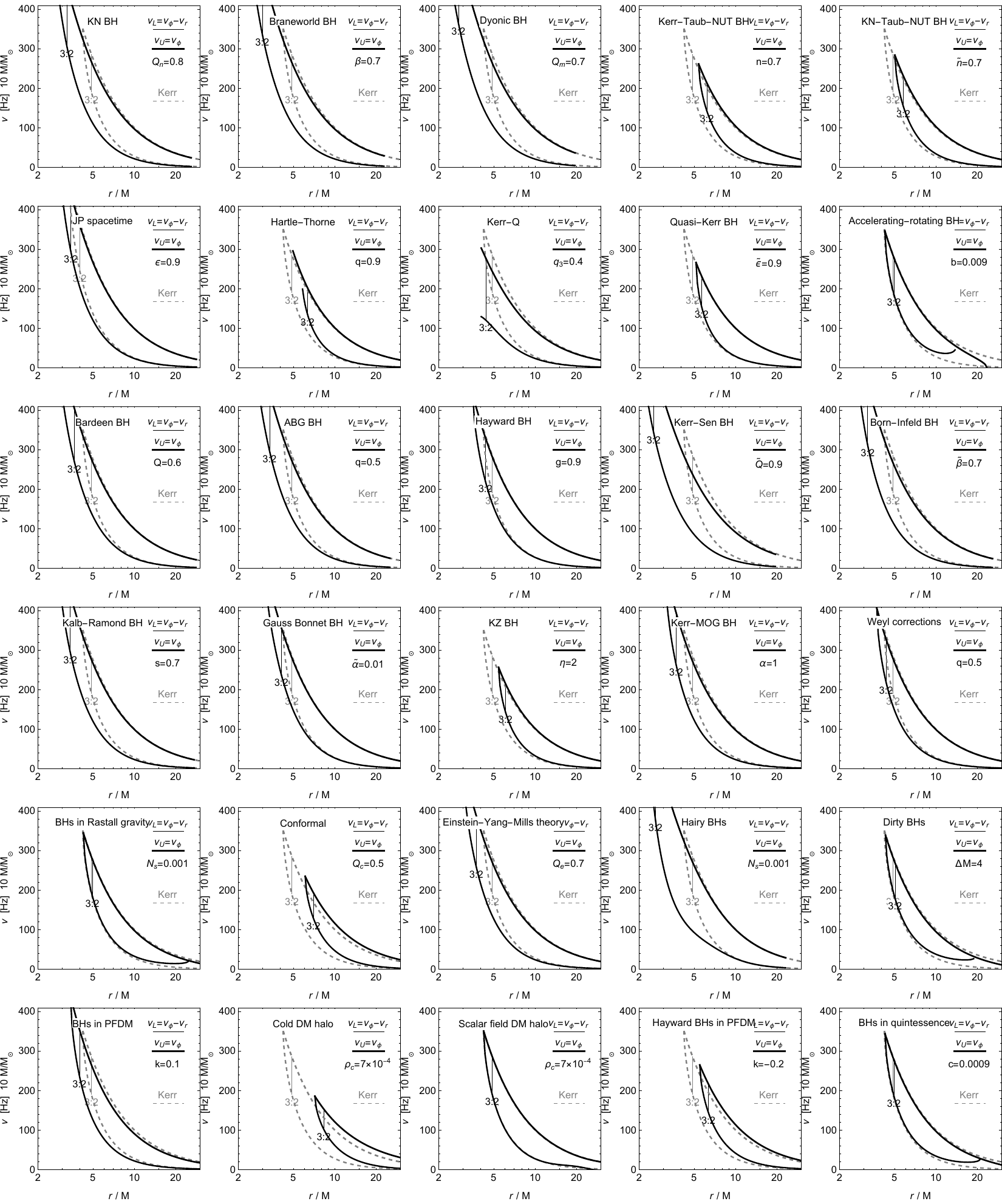}
\caption{Radial profiles of lower $\nu_{\rm L}(r)$ and upper $\nu_{\rm U}(r)$ frequencies for RP HF QPOs model in the background of many different stationary, axisymmetric and asymptotically flat spacetimes. We compare the Kerr BH limit with the well-motivated deviations from the classical Kerr BH solution. The frequencies are plotted for classical Kerr BH as gray curves, while black curves are for non-Kerr BHs. The BH spin is taken $a=0.5$ for all spacetimes. The position $\nu_{\rm U}:\nu_{\rm L}=3:2$ of resonant radii $r_{3:2}$ has also been shown by vertical lines.
\label{RP-32ratio}}
\end{figure*}
%-------------------------------------------------------------------------%

%-------------------------------------------------------------------------%
\begin{figure*}
\centering
\includegraphics[width=\hsize]{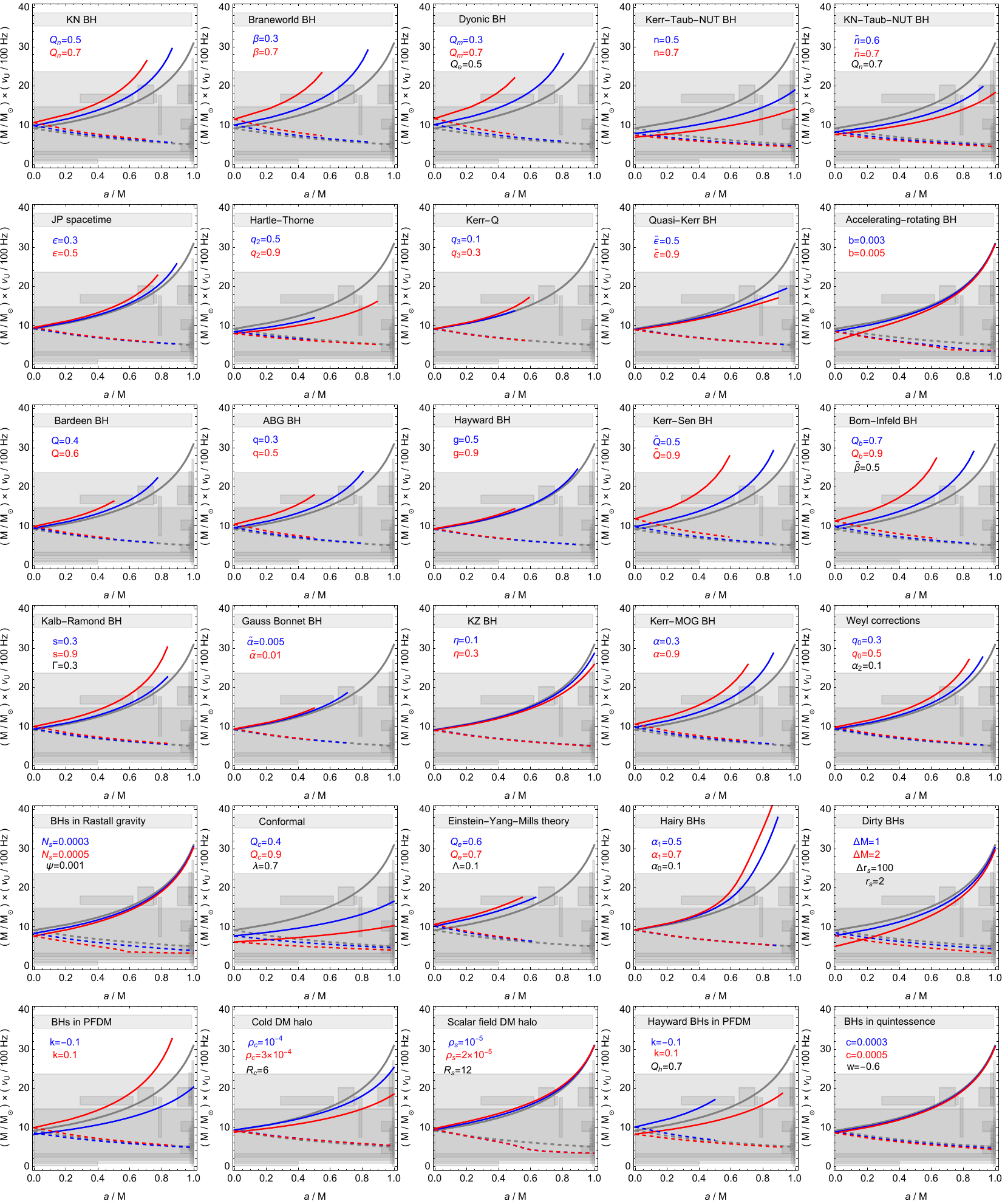}
\caption{Fitting of observational data of QPOs observed in the well-known microquasars as well as AGNs in the framework of the ER model of geodesic oscillations of Keplerian disks modified for the epicyclic oscillations of neutral test particles orbiting many different kinds of rotating BHs. The BH spin $a/M$ is given on the horizontal axis, while on the vertical axis, we give BH mass $M$ divided by the observed upper HF QPOs frequency $\nu_{\rm U}$. Gray boxes show the values of BH spin $a$ and mass $M$, estimated from the observations, see Tab.~\ref{tab1}. Solid curves represent the upper test particle frequency at $r_{3:2}$ resonant radii $f_{\rm U} = \nu_r(r_{3:2},a)$ as the function of BH spin $a$ for co-rotating particles, while dashed curves are plotted for contra-rotating particles. Gray curves depict the Kerr limit. 
\label{ER_fitsfig}}
\end{figure*}
%-------------------------------------------------------------------------%

%-------------------------------------------------------------------------%
\begin{figure*}
\centering
\includegraphics[width=\hsize]{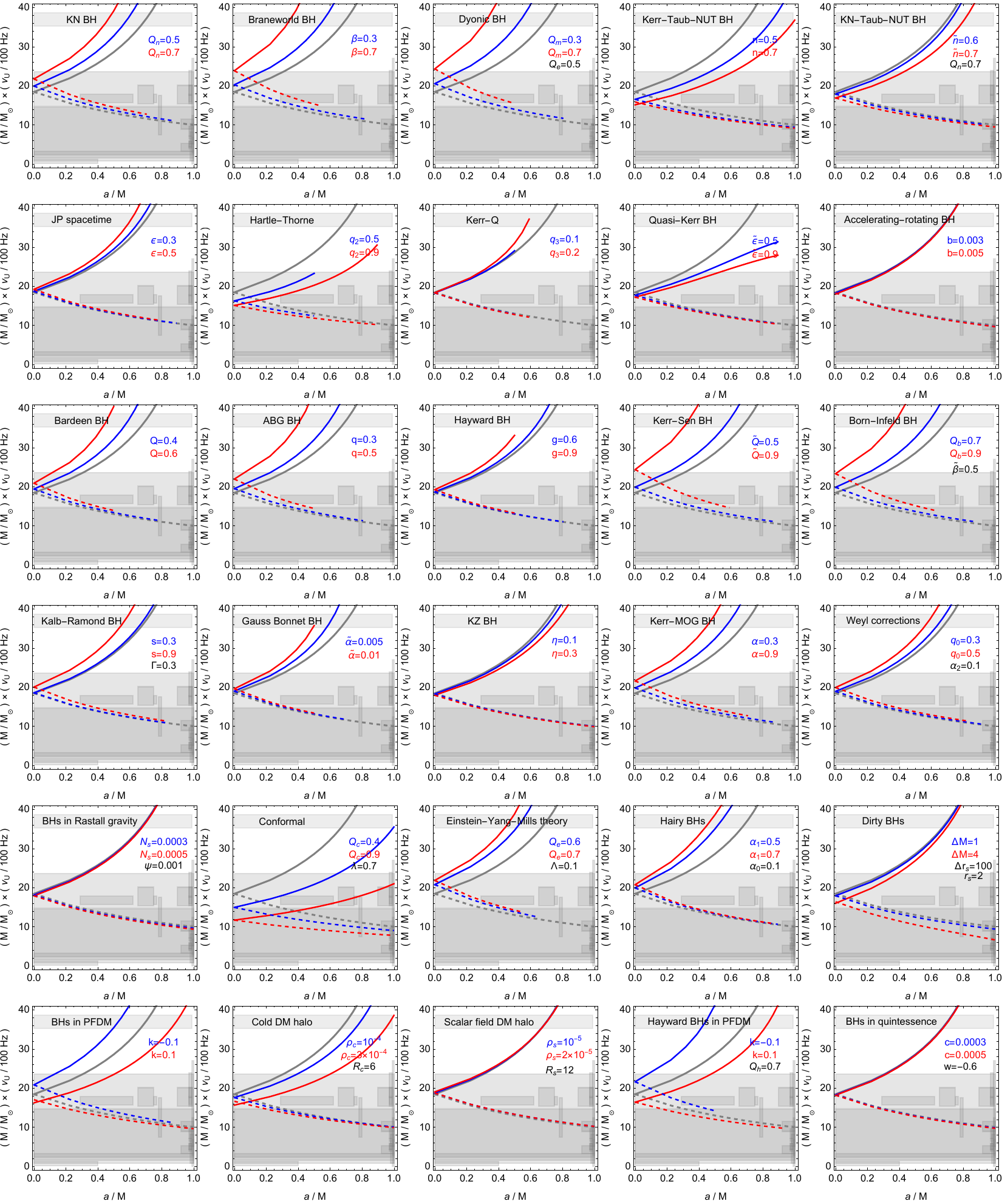}
\caption{Fitting of observational data of QPOs observed in the well-known microquasars as well as AGNs in the framework of the RP model of geodesic oscillations of Keplerian disks modified for the epicyclic oscillations of neutral test particles orbiting many different kinds of rotating BHs. The BH spin $a/M$ is given on the horizontal axis, while on the vertical axis, we give BH mass $M$ divided by the observed upper HF QPOs frequency $\nu_{\rm U}$. Gray boxes show the values of BH spin $a$ and mass $M$, estimated from the observations, see Tab.~\ref{tab1}. Solid curves represent the upper test particle frequency at $r_{3:2}$ resonant radii $f_{\rm U} = \nu_r(r_{3:2},a)$ as the function of BH spin $a$ for co-rotating particles, while dashed curves are plotted for contra-rotating particles. Gray curves depict the Kerr limit. 
\label{RP_fits}}
\end{figure*}
%-------------------------------------------------------------------------%

%%%%%%%%%%%%%%%%%%%%%%%%%%%%%%%%%%%%%%%%%%%%%%%%%%%%%%%%%%%%%%%%%%%%%%%%%%%
\section{Orbital models of HF QPOs} 
%%%%%%%%%%%%%%%%%%%%%%%%%%%%%%%%%%%%%%%%%%%%%%%%%%%%%%%%%%%%%%%%%%%%%%%%%%%

The QPOs in X-ray flux have long been detected in the stellar-mass BH binaries and are regarded as an efficient test of models of strong gravity. The oscillations of neutral particles around circular orbits, explored in the previous section, propose an interesting astrophysical application, related to HF QPOs detected in Galactic low mass X-ray binaries containing BHs \citep{Rem:2005:ASTRN:,Bell-etal:2012:MNRAS} or neutron~stars \citep{Bar-Oli-Mil:2005:MONNR:}. There are two AGNs, i.e., 2XMM J123+1106, and KIC 9650712 in which LF QPOs candidates have been declared. The LF QPOs identification was based on mass scaling of frequency assuming a linear relation between the mass of BH and LF QPOs period, yielding consistent masses with other independent estimates, as well as low coherence and high fractional variability consistent with LF QPOs. Since LF QPOs are not likely to be caused by the same process as HF QPOs, hence, we do not include them in our analysis.

Some HF QPOs have been identified in sources where there is no spin or mass determination. The first AGN QPO observation in RE J1034+396 has a high coherence and a measured BH mass consistent with the known mass-frequency scaling relation for HF QPOs in XRBs. As in several of the other cases, assuming the detected QPO frequency was an LF QPO results in mass upper limits that are inconsistent with the BH mass measurements from independent sources and are often debatable low for AGN  $(\sim 10^{5} M_{\odot})$.

We have selected four microquasars, i.e., GRO~1655-40, XTE~1550-564, XTE~1650-500, and GRS~1915+105, and many AGNs presented in Fig. \ref{Quasars} and Tab.~\ref{tab1}. The masses $M$ and spins $a$ of the central objects for all four microquasars have been estimated, but there are some AGNs for which spin estimates can not be found in literature, while the mass estimates exist with large errors, see shaded regions in the middle plot of Fig.~\ref{Quasars}. There also exist some AGNs for which only a lower limit of spin is known.     

% HERE SgrA flares vs QPOs

Our Galactic centre, Sgr~A* source, has attracted large attention recently by \cite{Grav-Coll:2018:A&A, EHTSC:2022:ApJ:}. In our previous work, we have been discussing the hot spot dynamics as a model for observed flares \citep{Shahzadi-et-al:2022:EPJC:} testing alternative theories of gravity. The QPOs for Sgr~A* source has been already reported in \citep{Asch:2004:AAP,Tor-etal:2005:ASTRA:}, but there is a growing concern if such data are not only red noise reported as a signal. The light curves duration to QPOs period ratio is much much larger for microquasars than for AGNs, hence the noise contamination is a smaller problem for microquasars as well. For AGNs or Sgr~A* sources, the data must be examined very carefully. It is also interesting that reported HF QPOs for Sgr~A* are at 0.8 mHz (1.4 mHz), see Tab.~\ref{tab1}. The observed flares around Sgr~A* have periods of 30-70 min, corresponding to Sgr~A* flares frequency 0.2-0.5 mHz, and are not that far from QPOs frequency 0.8 mHz (1.4 mHz). Hence, Sgr~A* flares and QPOs could be related although they are observed in different energy bands (IR vs. X-ray) \citep{Por-Miz-You:2021:MNRAS,Razieh-etal:2023:Gal:}.

The twin HF QPOs models involving the orbital motion of matter around BH can be classified into four classes: the hot spot models (the RP model and its variations \citep{Ste-Vie:1999:PHYSRL:,Stu-Kot-Tor:2013:ASTRA:}, the tidal precession model \citep{Kos-etal:2009:ASTRA:}), resonance models \citep{Tor-etal:2005:ASTRA:,Stu-Kot-Tor:2011:ASTRA:,Stu-etal:2015:ACTA:} and disko-seismic models \citep{Rez-etal:2003:MNRAS:,Mon-Zan:2012:MNRAS:}. These models are applied to match the twin HF QPOs for the microquasar GRO J1655-40 in \citep{Stu-Kol:2016:ASTRA:}. The models can also be applied for intermediate massive BHs \citep{Stu-Kol:2015:MNRAS:}. In this article, we will consider two well-known HF QPOs models, i.e., ER and RP models. One way of demonstrating this QPO phenomenon in the microquasars or AGNs is the formation of the oscillatory test particle motion around magnetized Kerr BHs \citep{Kol-Stu-Tur:2015:CLAQG:,Tur-Stu-Kol:2016:PHYSR4:,Kol-Tur-Stu:2017:EPJC:} or spinning test particles around Kerr BH \citep{Sha-et-al:2021:EPJC:}. Here, we explore the other modification of the strong gravity due to the modifications of classical Kerr BH.

%%%%%%%%%%%%%%%%%%%%%%%%%%%%%%%%%%%%%%%%%%%%%%%%%%%%%%%%%%%%%%%%%%%%%%%%%%%
\subsection{The ER model} 
%%%%%%%%%%%%%%%%%%%%%%%%%%%%%%%%%%%%%%%%%%%%%%%%%%%%%%%%%%%%%%%%%%%%%%%%%%%

The ER model is one of the popular variants of the resonance models characterized by the epicyclic parametric resonance model that supposes a $3:2$ non-linear resonance between the axisymmetric radial and latitudinal epicyclic modes of accretion disc oscillations \citep{Abr-Klu:2001:AA:,Tor-etal:2005:ASTRA:}. All the estimates obtained so far that were associated with the ER model assume a geodesic approximation of the accreted fluid motion. In this approximation, the two observable resonant frequencies are illustrated by the radial $\nu_{r}$ and latitudinal $\nu_{\theta}$ epicyclic frequencies of test particle motion and given by
\beq
\nu_{\rm L} = \nu_{r}, \quad \nu_{\rm U} = \nu_{\theta},
\eeq
where $\nu_{\rm L}$ and $\nu_{\rm U}$ denote the upper and lower frequencies of the twin peak with a frequency ratio close to $3:2$. In the ER model, the oscillating torus is supposed to be radiating uniformly. A sufficiently large inhomogeneity on the radiating torus, which orbits with the frequency $\nu_{\phi}$, enables the introduction of the nodal frequency related to this inhomogeneity \citep{Stu-Kol:2016:ASTRA:,Stu-Kol:2016:APJ:}. In more general accretion flows, non-geodesic effects associated for instance with magnetic fields or other forces may affect the characteristics of the considered oscillation modes and hence influence the spin predictions \citep{Kol-Stu-Tur:2015:CLAQG:,Kol-Tur-Stu:2017:EPJC:}. The radial profiles of upper $\nu_{\rm U}$ and lower $\nu_{\rm L}$ frequencies of the ER model for neutral test particles in the background of Kerr BH are presented in Fig.~\ref{Kerr_fits}, while the fundamental frequencies for a wide class of stationary, axisymmetric and asymptotically flat spacetimes are shown in Fig.~\ref{ER-32ratio}. We compare the motion of particles orbiting Kerr BH with the particle motion around many different non-Kerr BHs and examine the deviations from the Kerr limit. The radial profiles of frequencies of particles moving around non-Kerr BHs can be observed differently than the case of Kerr BH when the particle moves close to the BH but the radial profiles coincide as the particles move far away from the BHs. However, a clear deviation can be seen for the case of Kerr-Sen BHs in heterotic string theory and the BHs surrounded by cold DM halo.

%%%%%%%%%%%%%%%%%%%%%%%%%%%%%%%%%%%%%%%%%%%%%%%%%%%%%%%%%%%%%%%%%%%%%%%%%%%
\subsection{The RP model} 
%%%%%%%%%%%%%%%%%%%%%%%%%%%%%%%%%%%%%%%%%%%%%%%%%%%%%%%%%%%%%%%%%%%%%%%%%%%

The RP model was originally proposed to describe QPOs in low-mass X-ray binaries with a neutron star and then extended to systems with stellar-mass BHs \citep{Sta-Vie:1998:ApJ:}. It does not explain the origin of the QPOs, but it relates the observed frequencies of the QPOs with the fundamental frequencies of the background spacetime. In this model, the lower of the twin frequencies is described with the periastron precession frequency, while the upper one is described by the orbital frequency, given by
\beq
\nu_{\rm L} = \nu_{\phi} - \nu_{r}, \quad \nu_{\rm U} = \nu_{\phi},
\eeq
where upper $\nu_{\rm L}$ and lower $\nu_{\rm U}$ frequencies are the frequencies of the twin peak with a frequency ratio close to $3:2$. The radial profiles of upper $\nu_{\rm U}$ and lower $\nu_{\rm L}$ frequencies of the RP model for neutral test particles in the background of Kerr BH are presented in Fig.~\ref{Kerr_fits}, while the fundamental frequencies for many different rotating non-Kerr metrics are shown in Fig.~\ref{RP-32ratio}. The comparison of the motion of particles orbiting Kerr BH with the particle motion around many different non-Kerr metrics has been depicted. A clear deviation can be seen for the case of Kerr-Sen BHs in heterotic string theory and the BHs surrounded by cold DM halo.

%%%%%%%%%%%%%%%%%%%%%%%%%%%%%%%%%%%%%%%%%%%%%%%%%%%%%%%%%%%%%%%%%%%%%%%%%%%
\subsection{Resonant radii and the fitting technique} 
%%%%%%%%%%%%%%%%%%%%%%%%%%%%%%%%%%%%%%%%%%%%%%%%%%%%%%%%%%%%%%%%%%%%%%%%%%%

The HF QPOs observed in BH microquasars are generally stable and appear in a pair of two peaks with upper $\nu_{\rm U}$ and lower $\nu_{\rm L}$ frequencies in the timing spectra. The original resonance models induced the non-linear coupling between the radial and vertical epicyclic frequencies, or between the radial and orbital epicyclic frequencies in the accretion disk. In the first case, a coupling was supposed when the frequencies were in a ratio of frequencies $\nu_{\rm U}: \nu_{\rm L}$ close to a fraction $3:2$, congruent to the peaks observed in a handful of X-ray power spectra known at that time. In order to understand this coupling, Hor{\'a}k \citep{Hor:2008:aap:} explored weak non-linear interactions between the epicyclic modes in slender tori and observed the strongest resonance between the axisymmetric modes when their frequencies were in a ratio $3:2$. The observations of this effect in different non-linear systems show the existence of the resonances between two modes of oscillations. In the case of geodesic QPO models, the observed frequencies can be expressed in terms of linear combinations of the particle fundamental frequencies $\nu_{r}$, $\nu_{\theta}$ and $\nu_{\phi}$. For neutral test particles moving around rotating non-Kerr BH, the upper and lower frequencies of HF QPOs are the functions of mass $M$ of the BH, rotation parameter $a$, the resonance position $r$, and some additional parameter $\alpha$ of BH as
\beq
 \nu_{\mathrm{U}} = \nu_{\mathrm{U}}(r,M,a, \alpha), \quad \nu_{\mathrm{L}} = \nu_{\mathrm{L}}(r,M,a, \alpha). \label{ffUL}
\eeq
The dependence of frequencies $\nu_{\mathrm{U}}$ and $\nu_{\mathrm{L}}$ on the BH mass $M$, spin $a$ and additional parameter of BH is complicated and hidden inside $\Omega_{r}, \Omega_{\theta}, \Omega_{\phi}$ functions, as given by Eq. (\ref{nu_rel}). In order to fit the frequencies observed in HF QPOs with the BH parameters, one needs first to calculate the so-called resonant radii $r_{3:2}$, given by
\beq
\nu_{\mathrm{U}}(r_{3:2}):\nu_{\mathrm{L}}(r_{3:2})=3:2 \label{rezrad}.
\eeq
For given values of spin parameter $a$, the resonant radii $r_{3:2}$ are usually represented by a numerical solution of higher order polynomial in $r$. Since Eq. (\ref{rezrad}) is independent of the mass of BH, thus the resonant radius solution also has no explicit dependence on the BH mass. In order to find the resonant radii, we use the technique proposed in \citep{Stu-Kot-Tor:2013:ASTRA:}. Substituting the resonance radius into Eq. (\ref{ffUL}), we obtain the frequencies $\nu_{\mathrm{U}}$ and $\nu_{\mathrm{L}}$ in terms of the BH mass $M$, spin $a$ and parameter $\alpha$. Now, we can compare the calculated frequencies $\nu_{\mathrm{U}}$ and $\nu_{\mathrm{L}}$ with observed HF QPOs frequencies $f_{\mathrm{U}}$ and $f_{\mathrm{L}}$, see Tab.~$1$. If one assumes that observed frequencies are exactly in $3:2$ ratio, $f_{\mathrm{U}}:f_{\mathrm{L}}=3:2$, then the following equations
\beq
f_{\mathrm{U}} = \nu_{\mathrm{U}}(r,M,a, \alpha), \quad f_{\mathrm{L}} = \nu_{\mathrm{L}}(r,M,a, \alpha),\label{EQ}
\eeq
are equivalent with
\beq
f_{\mathrm{U}} = \nu_{\mathrm{U}}(r_{3:2},M,a, \alpha).
\eeq
The position of resonant radii $r_{3:2}$ for HF QPO ER and RP models for many different non-Kerr spacetimes is depicted in Figs.~\ref{ER-32ratio} and \ref{RP-32ratio}, respectively. The positions of resonant radii $r_{3:2}$ for the case of the HF QPO RP model exist closer to the BH as compared to the ER model. Substituting the resonance radius $r_{3:2}$ into the Eq. (\ref{ffUL}), we obtain the frequencies $\nu_{\mathrm{U}}$ and $\nu_{\mathrm{L}}$ in terms of the BH mass, rotation $a$ and additional parameter of BH. Now, we can compare the calculated frequencies $\nu_{\mathrm{U}}$ and $\nu_{\mathrm{L}}$ with the observed HF QPOs frequencies $f_{\mathrm{U}}$ and $f_{\mathrm{L}}$, respectively, see Tab.~\ref{tab1}. The calculated upper frequency at resonant radii $r_{3:2}$ as a function of BH spin $\nu_{\mathrm{U}}(a)$ has been used to fit the observed BH spin and mass data in Figs.~\ref{ER_fitsfig} and \ref{RP_fits} for ER and RP QPO models, respectively. For the HF QPO ER model, the frequencies for contra-rotating particles do not fit any microquasars, but the co-rotating particles fit all microquasars very well. The best fitting of observational data of QPOs in the background of ER model can be seen in the case of BHs surrounded by PFDM. However, in the case of the RP model, the frequencies for contra-rotating particles fit only a few microquasars and AGNs, but the co-rotating particles do not fit any microquasars or AGNs sources except rotating regular BHs in conformal massive gravity. We observe that all considered BHs fit the observational data of QPOs around Sgr~A* for the RP model, however, we don't see any fitting for ER model except hairy BHs.

%%%%%%%%%%%%%%%%%%%%%%%%%%%%%%%%%%%%%%%%%%%%%%%%%%%%%%%%%%%%%%%%%%%%%%%%%%%
\section{Conclusions} \label{kecy}
%%%%%%%%%%%%%%%%%%%%%%%%%%%%%%%%%%%%%%%%%%%%%%%%%%%%%%%%%%%%%%%%%%%%%%%%%%%

The astrophysical BH candidates are thought to be the Kerr BHs predicted by GR, but the actual nature of these objects has still to be confirmed. In order to test the Kerr nature of these objects, one needs to probe the geometry of the spacetime around them and check if the observations are consistent with the predictions of the Kerr metric, which can be achieved in a number of ways. Many observations, for instance, HF QPOs observed in the X-ray flux of some stellar-mass BHs \citep{Smi-Tan-Wag:2021:APJ:}, hot-spot data of three flares observed by the GRAVITY instrument \citep{GRAVITY:2018:A&A:}, and BH shadow of Sgr~A* observed by Event Horizon Telescope \citep{EHT:2022:ApJ:} can help us to determine the nature of spacetime. Moreover, continuum-fitting \citep{McClintock-etal:2011:CQGra:,Bambi-etal:2016:CQGra:} and the iron line \citep{Fabian-etal:1995:MNRAS:,Reynolds:2003:PhR:} methods could also be used to test the Kerr BH paradigm.  

We consider a wide range (30 in total) of well-motivated deviations from classical Kerr BH and explore the neutral test particle fundamental frequencies of small harmonic oscillations around circular orbits. Recently, we have used the same set of stationary, axisymmetric, and asymptotically flat spacetimes, and estimated the restrictions of spacetime parameters with the help of hot-spot data of three flares observed at Sgr~A* by GRAVITY instrument \citep{Shahzadi-et-al:2022:EPJC:}. The constraints of parameters for many different spherically symmetric spacetimes have been obtained using BH shadow of Sgr~A* observed by Event Horizon Telescope in \citep{Sunny-etal:2023CQGra:}. 

The objective of this work is not to test a specific theoretical model and to determine or constrain its parameters. Instead, our aim is to review a class of stationary, axisymmetric, and asymptotically flat spacetimes and investigate the possible deviations from the Kerr geometry of the spacetime around astrophysical BHs. The HF QPOs observed in some stellar-mass BH candidates may be used to test the Kerr-nature of these objects and to confirm, or rule out, the Kerr BH hypothesis. These QPO frequencies remain almost constant, indicating that they are not influenced by the characteristics of the accretion flow but rather by the spacetime geometry itself. We explored the perturbations of circular orbits and investigated the radial profiles of lower and upper frequencies for HF QPOs ER and RP models in the background of well-motivated deviations from classical Kerr BH. In the present work, we are demonstrating the general behaviour of fundamental (orbital) frequencies for various different axially spacetimes, demonstrating their differences. Our work could serve as a guidepost for any future detailed works on any chosen spacetime and our review article could help to select BH spacetime with desired orbital frequency behavior. For example, a clear deviation can be observed in the case of Kerr-Sen BHs in heterotic string theory and the BHs in PFDM. It is concluded that the frequencies of particles moving around non-Kerr BHs coincide with that of Kerr BH when the particles move away from the BHs.  
%The HF QPOs are an excellent tool to study the geometry of the spacetime around steller-mass BH candidates and to measure the fundamental properties of these objects. 

% regarding the resonant radii in Fig. 3 & 4
The positions of resonant radii, for example, $r_{3:2}$ or $r_{1:2}$, play a crucial role in understanding the behavior of test particle motion around BHs. Resonances can significantly alter the dynamics of test particles around BHs and highlight the deviations from the standard Kerr spacetime to non-Kerr spacetime \citep{Apostolatos-etal:2009:PhRvL:,Georgios-etal:2010:PhRvD:}. We examined the positions $\nu_{\rm U}:\nu_{\rm L}=3:2$ of resonant radii $r_{3:2}$ for both HF QPO ER and RP models in the background of several non-Kerr spacetimes and observed the resonant radii closer to the BH for the case of RP model as compared to the ER model. It is noted that for most of the considered non-Kerr spacetimes, the resonant radii are positioned at smaller radii than those predicted by the standard Kerr spacetime.

We have presented a generalized formalism for fitting observational data of QPOs observed in the well-known microquasars, AGNs, and Sgr~A* within the framework of the ER and RP models of geodesic oscillations of Keplerian disks modified for the epicyclic oscillations of both co-rotating and contra-rotating particles orbiting various types of rotating BHs. Our investigation primarily focuses on reviewing a specific class of stationary, axisymmetric, and asymptotically flat spacetimes, and comparing the observational data of HF QPOs with the fittings obtained from these spacetimes. It is found that the best fit for observational data in the background of ER and RP models can be observed for BHs surrounded by PFDM, and the regular BHs in conformal massive gravity, respectively. For the case of the ER model, only Hairy BHs are capable of fitting the observational data for Sgr~A* among all the considered BH spacetimes. However, in the case of the RP model, all BH spacetimes that have been examined are able to fit the observational data for Sgr~A*, except Hartle-Thorne spacetime and Quasi-Kerr BHs. It is worth mentioning that the best fit can be obtained for the microquasar GROJ1655-40 in the case of the ER HF QPO model. However, for the RP model, the best fit can be obtained for XTE J1650-500. As we can see there is currently no spacetime that can explain (fit) data for all microquasar or AGNs sources, especially it seems that it is quite hard to fit data for Sw~J164449.3+573451, ASASSN-14li, and Sgr~A* sources using neutral particle orbital frequencies. This could mean that more complex QPO models like with magnetic field influence \citep{Kol-Tur-Stu:2017:EPJC:} must be included or that the observed frequencies are not correct HF QPOs.
% regarding the fits in Fig. 5 & 6
%\ms{The parameters of BH spacetimes such as the spin parameter or charge parameter are very important to determine the properties of BH spacetimes. For example, the position of ISCOs or the horizon of BH is highly sensitive to the spin parameter. In order to prevent the occurrence of BH naked singularity, the spin parameter $a=0.5$ has been used for all considered BH spacetimes. A better fitting can be obtained by using the BH spin parameter, $0.5<a<1$, this requires the use of small values of other BH parameters corresponding to considered non-Kerr BH spacetime, which closely resembles the Kerr case. However, a clear deviation from Kerr spacetime can be observed if the large values of BH parameters are employed, but this requires the use of a small spin parameter, $0<a<0.5$. As a result, the curves of frequencies end before reaching the high spin ($a=1$), and the best fitting can not be achieved.}

%\ms{We have compared our obtained results with our previous study on the motion of hot-spot around a wide class of stationary, axisymmetric, and asymptotically flat spacetimes, fitting the observed positions and periods of three flares observed at Sgr~A* detected by GRAVITY instrument \citep{Shahzadi-et-al:2022:EPJC:}. In both studies, the BHs surrounded by PFDM are the most promising candidates that are demonstrating the best tendency to fit the observational data among all the considered BH spacetimes.}

%%%%%%%%%%%%%%%%%%%%%%%%%%%%%%%%%%%%%%%%%%%%%%%%%%
\section*{Data Availability}
%%%%%%%%%%%%%%%%%%%%%%%%%%%%%%%%%%%%%%%%%%%%%%%%%%

The observational data associated with the manuscript has been presented in Fig.~\ref{Quasars}, and Tab.~\ref{tab1}, there is no additional data for this article.
 
%%%%%%%%%%%%%%%%%%%%%%%%%%%%%%%%%%%%%%%%%%%%%%%%%%%%%%%%%%%%%%%%%%%%%%%%%%%
\section*{Acknowledgments}
%%%%%%%%%%%%%%%%%%%%%%%%%%%%%%%%%%%%%%%%%%%%%%%%%%%%%%%%%%%%%%%%%%%%%%%%%%%

This work is supported by the Research Centre for Theoretical Physics and Astrophysics, Institute of Physics, Silesian University in Opava, and Czech Science Foundation Grant No.~\mbox{23-07043S}.
%The authors MS, MK, and ZS would like to express their acknowledgment for the Research Centre for Theoretical Physics and Astrophysics, Institute of Physics, Silesian University in Opava, and Czech Science Foundation Grant No. \mbox{23-07043S}.

\def\prc{Phys. Rev. C}
\def\pre{Phys. Rev. E}
\def\prd{Phys. Rev. D}
\def\jcap{Journal of Cosmology and Astroparticle Physics}
\def\apss{Astrophysics and Space Science}
\def\mnras{Monthly Notices of the Royal Astronomical Society}
\def\apj{The Astrophysical Journal}
\def\aap{Astronomy and Astrophysics}
\def\actaa{Acta Astronomica}
\def\pasj{Publications of the Astronomical Society of Japan}
\def\apjl{Astrophysical Journal Letters}
\def\pasa{Publications Astronomical Society of Australia}
\def\nat{Nature}
\def\physrep{Physics Reports}
\def\araa{Annual Review of Astronomy and Astrophysics}
\def\apjs{The Astrophysical Journal Supplement}
\def\prl{Physical Review Letters
}
\def\aapr{The Astronomy and Astrophysics Review}
\def\procspie{Proceedings of the SPIE}

\bibliography{reference}
\bibliographystyle{unsrt}

\end{document}